\documentclass[reqno]{amsart}
\pdfoutput=1
\usepackage{graphicx}
\usepackage{amsmath,amssymb,graphics,epsfig,color}
\usepackage[mathscr]{euscript}
\theoremstyle{plain}

\theoremstyle{definition}

\theoremstyle{remark}

\numberwithin{equation}{section}
\numberwithin{theorem}{section}

\renewcommand{\epsilon}{\varepsilon}
\renewcommand{\tilde}{\widetilde}
\renewcommand{\hat}{\widehat}

\definecolor{light}{gray}{.9}

\title[Strong particles]{Gradient structure and transport coefficients for strong particles }

\author[D.\ Gabrielli]{Davide Gabrielli}
\address{Davide Gabrielli \hfill\break \indent
  DISIM, Universit\`a dell'Aquila
  \hfill\break\indent
  Via Vetoio,   67100 Coppito, L'Aquila, Italy}
\email{gabriell@univaq.it}

\author[P. L. Krapivsky]{P. L. Krapivsky}
\address{P. L. Krapivsky \hfill\break\indent
Department of Physics, Boston University,
\hfill\break\indent
Boston, Massachusetts 02215, USA}
\email{paulk@physics.bu.edu}

\begin{document}

\begin{abstract}
We introduce and study a simple and natural class of solvable stochastic lattice gases.
This is the class of \emph{Strong Particles}. The name is due to the fact that when they try to jump to an occupied site they succeed pushing away a pile of particles.  For this class of models we explicitly compute the transport coefficients. We also discuss some generalizations and the relations with other classes of solvable models.

\bigskip


\smallskip

\end{abstract}

\maketitle
\thispagestyle{empty}

\section{Introduction}

Stochastic interacting particle systems are very important as toy models in statistical mechanics \cite{Spohn,KL99,S00,BE07}. In this class of models, particles evolve on a lattice under a stochastic Markovian evolution. One of the most well-known interacting particle system is the simple exclusion process (SEP). In this interacting particle system each lattice site is either occupied by a particle or empty, and particles undergo hopping to neighboring sites according to independent random walks. Hops to empty sites are allowed, while hops to occupied sites are forbidden. This is one of the simplest models \cite{Spohn,KL99,S00,BE07,D07,MFTReview,CKZ} for which the hydrodynamic collective behavior of the system is well understood with an explicit knowledge of the transport coefficients. Here we investigate a class of models which are almost as tractable as the SEP. For these models we write down an explicit hydrodynamic equation and compute the transport coefficients. This is due to the fact that, as in the case of SEP, we explicitly know the invariant measure and moreover the models are of \emph{gradient type}.

In general it is a difficult task to construct a gradient stochastic lattice gas that is reversible and for which the invariant measure is explicitly known. This is what happens for our class of models. We call them a \emph{Strong Particles} (SP) process. Particles evolve stochastically on a regular lattice satisfying an exclusion rule. The particles are strong and when they try to jump to an occupied site they succeed pushing away a pile of particles. The rate of jump may depend on the number of particles pushed away. The microscopic gradient structure for the SP is not as simple and evident as in the case of the SEP and we discuss it in detail. We focus here on the one-dimensional case, generalizations to higher dimensions are possible.

The basic property of the models that implies the validity of a gradient condition is the following. For any configuration of particles it is possible to define an associated instantaneous current across each bond of the lattice. For the SP, the total sum of the currents on all the bonds of each cluster of particles vanishes.  This fact allows us to construct a function whose gradient is the instantaneous current. This is done fixing at zero its value on each empty site. Thus essentially we fix as the reference value zero the height of the ``sea" of the empty sites.

Since the particles are indistinguishable this dynamics is equivalent to the long jump exclusion dynamics discussed in \cite{SOC:sing, sing_diff} that we call leap frog dynamics. Generalizations of these two dynamics to the case of evolving hard rods which we also study in this paper are different, and the generalization relying on strong particles appears more physical; henceforth we employ the strong particles interpretation. Our formula \eqref{kap} is equivalent to the formula for the diffusion coefficient in \cite{sing_diff}. Our computations elucidate the discrete geometric structure that is behind the solvability of the model. (We call solvable a model for which the macroscopic collective behavior can be completely understood.) Further, we demonstrate the validity of the Einstein relation and we extend and generalize the construction to different frameworks.  We also describe generalizations of the SP models to the situations when the invariant measure is not product, we consider the cases when the exclusion constraint is relaxed, and we discuss strong extended objets (hard rods). We give explicit formulas for the transport coefficients in these more complicated models in a number of special situations. The behaviors of the tagged particle in the SP and in the leap frog dynamics are different. As far as we known the limiting behavior of the tagged particle for a SP model has not been investigated before and it is an interesting open problem (see \cite{Pab} for a discussion of problems of this type). In the case of hard rods the strong dynamics and the leap frog dynamics give two different evolutions of the mass of the system.

We also discuss the relation between solvability of the SP and solvability of other stochastic lattice gases. In particular we show how to deduce the transport coefficients for the Hammersley process \cite{Pab} and for an exclusion process with avalanches \cite{EPA} starting from the transport coefficients for the SP. Several other models are similar or strictly related to the SP, see e.g. \cite{FS,DLSS, DLSS2,SRB,SS,BF_08} where models with asymmetric hopping have been studied. We assume that the hopping rules are symmetric (and briefly discuss weakly asymmetric models).

\smallskip

Throughout this paper we analyze one-dimensional systems with periodic boundary conditions. In a couple of Sections we consider different boundary conditions; in these cases we will warn the reader. Sometimes we use higher-dimensional notation since the arguments can be straightforwardly generalized to higher dimensions, but we focus on the one-dimensional setting.

\section{Macroscopic description}

We first outline the general properties of the large scale hydrodynamic behavior of lattice gases
without explicit references to specific models (see \cite{Spohn,KL99,D07,MFTReview} for more details). For a large class of diffusive lattice gases, the only relevant hydrodynamic variable is the density field $\rho(x,t)$ that evolves according to a non-linear diffusion equation
\begin{equation}
\label{DE}
\partial_t \rho = \nabla \cdot \left[D(\rho)\, \nabla \rho\right].
\end{equation}
Here $\nabla$ is the gradient and $\nabla \cdot$ denotes the divergence.
The symmetric positive definite diffusion matrix $D(\rho)$ generically depends on the density and it encapsulates all microscopic rules underlying the macroscopic dynamics of the lattice gas.  In presence of a weak external field $E$ the hydrodynamic equation becomes
\begin{equation}
\label{DEE}
\partial_t \rho = \nabla \cdot \left[D(\rho)\, \nabla \rho\right]-\nabla\cdot \left[\sigma(\rho) E\right],
\end{equation}
where $\sigma(\rho)$ is a positive definite symmetric mobility matrix. It encapsulates all the information on the underlying stochastic microscopic lattice on the hydrodynamic response to the action of a weak external field.

The equilibrium origin of $D(\rho)$ and $\sigma(\rho)$ is emphasized by the Einstein relation between these two quantities and the free energy density $f(\rho)$:
\begin{equation}
\label{FDT}
2D(\rho)=\sigma(\rho)f''(\rho)\,.
\end{equation}
This relation states that for each value of $\rho$ the two matrices $D$ and $\sigma$ are proportional and
the proportionality coefficient is the second derivative of the density of the equilibrium free energy.

The free energy density is defined as follows. Consider the model on a $d$ dimensional lattice of side length $N$.
Assume that  in equilibrium its invariant measure is Gibbsian with a finite range Hamiltonian $H$. If we denote by $\eta$ a configuration of particles, we define the \emph{pressure} as (see \cite{Lanf})
\begin{equation}\label{pres}
p(\lambda)=\lim_{N\to +\infty}\frac{1}{N^d}\log \sum_{\eta} e^{-H(\eta)}e^{\lambda\sum_i\eta(i)}\,,
\end{equation}
where $\lambda\in \mathbb R$ is the chemical potential. The free energy density is then the Legendre transform of the pressure
\begin{equation}
\label{den-fr}
f(\rho)=\sup_{\lambda}\left\{\lambda\rho-p(\lambda)\right\}\,.
\end{equation}

This general framework holds in any dimensions, but we restrict ourselves to the one-dimensional situation. In particular, we will consider one-dimensional models for which the transport coefficients $D$ and $\sigma$ can be computed explicitly starting from the transition rates.

\section{Strong Particles }
\label{generators}

Strong Particles (SP) models are exclusion processes, so that each site is occupied by at most one particle. In the simplest versions of the SP dynamics, each hopping attempt is successful and the adjective {\em strong} emphasizes this feature.

Our models describe the evolution of stochastic indistinguishable particles, but their evolution can be better described and understood by labeling particles. Clearly the description of the stochastic evolution of the labeled particles is necessary if one wants to study the evolution of a tagged particle. More than one evolution of labeled particles can correspond to the same model of unlabeled particles. We will discuss this in detail in section \ref{LFMP}. We start considering SP with symmetric hopping rates. Then we will discuss generalizations  and deformations obtained by the introduction of an external field.

In the basic setting, SP perform independent simple random walks and each hopping attempt to a neighboring site is successful. If the selected site is occupied by another particle, the newly arriving particle merely pushes it away with the entire adjacent block of particles. Here are a few examples illustrating the hopping rules for SP in one dimension:
\begin{equation}
\label{SP:rules}
\begin{split}
\star\circ  &\longrightarrow \circ\star \\
\circ\star\bullet\bullet\bullet\bullet\circ &\longrightarrow \circ\circ\star\bullet\bullet\bullet\bullet\\
\bullet\circ\bullet\bullet\bullet\star\bullet &\longrightarrow \bullet\bullet\bullet\bullet\star\circ\bullet
\end{split}
\end{equation}
In these examples $\star$ denotes the particle which makes the hop, $\bullet$ denotes a site occupied by a particle, and $\circ$ denotes an empty site. In the basic symmetric version of the SP all hops happen with the same rate (which we set to unity).

The hopping rules for SP make sense on the finite hyper-cubic lattices in all dimensions. The extension of the SP process to other lattices is less natural. The definition and well-posedness of the model on the infinite lattice is a delicate issue that we will not discuss; see \cite{G,L} for a discussion of the problems involved.

We can define more formally our models as follows. Consider the ring with $N$ vertices. This is a graph with vertices $\Lambda_N:=\mathbb Z/N\mathbb Z$  and edges $E_N$ given by the ordered pairs of nearest neighbors vertices. We say also that $i+1$ is on the right of $i$ and $i$ is on the left of $i+1$. The configurations space is given by $\left\{0,1\right\}^{\Lambda_N}$ and an element is a configuration of particles $\eta$ such that $\eta(i)=1$ when there is a particle on the lattice site $i\in \Lambda_N$ and $\eta(i)=0$ when the lattice site is empty. Due to the exclusion rule we can not have more than one particle on each lattice site.

The stochastic evolution is a Markov jump dynamics on $\left\{0,1\right\}^{\Lambda_N}$. This is defined by
the transition rates $r(\eta,\eta')$ for a jump from a configuration $\eta$ to the configuration $\eta'$. The collection of the transition rates is codified in the generator of the process that fully describes the stochastic evolution. Given a function $f:\left\{0,1\right\}^{\Lambda_N}\to \mathbb R$ we have
\begin{equation}\label{gen}
\mathcal L_N f(\eta)=\sum_{\eta'}r(\eta,\eta')\big[f(\eta')-f(\eta)\big]\,.
\end{equation}
Let us now describe in detail the rates.
Given $i\in \Lambda_N$ we define
\begin{equation}\label{l+}
d^{\pm}_i(\eta):=\inf\left\{n \geq 0\,:\, \eta(i\pm n)=0\right\}\,.
\end{equation}
In the above formulas the sum is modulo $N$ and when $\eta(j)=1$ for all the lattice points $j$ then we define $d^{\pm}_i(\eta)=N$. If we are considering a system of particles in the infinite lattice $\mathbb Z$, the quantities $d^{\pm}_i(\eta)$ can assume the value $+\infty$. The integer numbers $d^{\pm}_i(\eta)$ represent the number of particles (including the particle at $i$ if any) that separate the particle at site $i$ respectively to the right and to the left from the first empty site. We have $d^{\pm}_i(\eta)=0$ when $\eta(i)=0$. We use also the shorthand $d^{\pm}_i$ when the configuration of particles considered is clear.

Let us introduce a family of operators $\sigma^{\pm}_{i}$ acting on configurations of particles. They are defined as follows. If $d^{\pm}_i(\eta)$ is equal to $0$ or $N$ then $\sigma^{\pm}_i\eta=\eta$. If instead $0<d^{\pm}_i<N$ then we define the configuration $\sigma^{\pm}_i\eta$ as
\begin{equation}\label{op+}
\left[\sigma^{\pm}_i\eta\right](j):=\left\{
\begin{array}{ll}
1 & \textrm{if}\ j=i\pm d_i^\pm\,,  \\
0 & \textrm{if} \ j=i\,,\\
\eta(j) & \textrm{otherwise}\,.
\end{array}
\right.
\end{equation}
These operators can be naturally interpreted in terms of the jumps of particles across the bonds of the lattice
when a strong particle jumps. The generator of the basic SP process is given by
\begin{equation}\label{gen-strong}
\mathcal L_Nf(\eta):=\sum_{i\in \Lambda_N}\sum_{s=\pm}\big[f\big(\sigma^{s}_i\eta\big)-f\big(\eta\big)\big]\,.
\end{equation}
We define a flow $q$ as a map from $E_N$ to $\mathbb R^+$. The quantity
$q(i,j)$ represents the amount of mass flowing through the edge $(i,j)$.
We define the divergence of a flow at site $i$ as
\begin{equation}\label{disc-div}
\nabla\cdot q(i)=\sum_j q(i,j)-\sum_jq(j,i)\,.
\end{equation}
Denote by $||q||:=\sum_{e\in E_N}q(e)$ the total mass flowing. We introduce the following elementary flows naturally associated to the operators $\sigma_i^{\pm}$:
\begin{equation}
\label{qflussi}
q^{\pm}_{i}(l,m):=\left\{
\begin{array}{ll}
1 & \textrm{if}\ (l,m)=\big(i\pm n ,i\pm(n+1)\big)\,, n \in \left\{0,\dots, d^{\pm}_i(\eta)-1\right\},\\
0 & \textrm{otherwise}.
\end{array}
\right.
\end{equation}
These are the flows corresponding to jumps of a strong particle at $i$. They depend on $\eta$ but to simplify notation we do not write explicitly such a dependence. We write $q^{\pm}_{i}[\eta]$ when we need to specify the configuration of particles.
It follows directly from the definition that
\begin{equation}
\label{disc-cont-q}
\sigma^{\pm}_{i}\eta=\eta-\nabla\cdot q^{\pm}_{i}\,.
\end{equation}

Consider now a function $r:\mathbb N\to \mathbb R^+$ satisfying $r(0)=0$. We can define generalized SP processes by
\begin{equation}
\label{gen-strong-k}
\mathcal L_N^rf(\eta):=\sum_{i\in \Lambda_N}\sum_{s=\pm}r\left(||q^{s}_i||\right)\big[f\big(\sigma^{s}_{i}\eta\big)-f\big(\eta\big)\big]\,.
\end{equation}
A natural description of the process in terms of labeled particles is as follows. As before, the particles interact by
hard core interaction and try to perform a nearest neighbor random walk.  The particles are strong so that
they can shift a pile of particles. If a particle must shift $n$ particles, we postulate that this occurs with rate $r(n+1)$; the notation reflects the fact that $n+1$ particles are moving, including the one that pushes the others. One simple choice is the step function
\begin{equation}
\label{strong-k}
r_k(n) =
\begin{cases}
1 & 1\leq n\leq k+1\\
0 & n>k+1
\end{cases}
\end{equation}
posing that each particle can push at most $k$ adjacent particles, and whenever the push is admissible it occurs with the same unit rate. The number $k$ quantifies the strength of the particles. We call this model the k-SP process. The 0-SP process is the simple exclusion process and the $\infty$-SP process is the basic SP process.

The values of $r(k)$ with $k>N$ are of course irrelevant for the dynamics on a ring with $N$ sites. When $N$ is becoming large, however, all the values of the function $r$ will become relevant and the limiting behavior will depend on all of them.

\section{Invariant measures and reversibility}

The models defined by \eqref{gen-strong-k} are conservative in the sense that they conserve the total number of particles. Under the assumption that $r(1)>0$, for any fixed number of particles the stochastic dynamics is irreducible so that any possible configuration can be reached from any other. Consequently, there is a one-parameter family of canonical invariant measures, one for each number of particles.

Indeed all the models are reversible with respect to the canonical uniform measures, i.e. for the fixed total number of particles $M$, the unique invariant measure for all the models is $\nu_{N,M}$ the probability measure that gives equal weight to all $\binom{N}{M}$ possible configurations with $M$ particles. This follows immediately from the validity of the detailed balance condition. Since any possible jump preserves the number of particles and since the uniform measures gives the same weight to configurations of particles with the same number of particles the detailed balance condition follows by the following simple observations. First we observe that if the system can jump from a configuration $\eta$ to a configuration $\eta'$ then it can perform also the reverse jump. Moreover both jumps correspond to the same number of particles moved and then they have the same rate, that is
\begin{equation}\label{det-bal}
r(\eta,\eta')=r(\eta',\eta)\,.
\end{equation}
Relation \eqref{det-bal} not only says that
for any $M$ $\nu_{N,M}$ is invariant for the dynamics but also that the processes are reversible.

By relation \eqref{det-bal} we deduce also that
$$
\nu_\rho(\eta)=\rho^{\sum_{i\in \Lambda_N}\eta(i)}(1-\rho)^{\sum_{i\in \Lambda_N}\left[1-\eta(i)\right]},
$$
a grand canonical Bernoulli product measure of parameter $\rho\in [0,1]$, is also invariant and reversible. More generally all the invariant measures are obtained as convex combinations of the canonical uniform ones.

By a direct computation using \eqref{pres} and \eqref{den-fr} the free energy density for a Bernoulli product measure $\nu_\rho$
is given by
\begin{equation}\label{fff}
f(\rho)=\rho\log\rho+(1-\rho)\log(1-\rho)\,.
\end{equation}

\section{Gradient condition}
\label{sezngrad}

Given an edge $(i,j)\in E_N$ and a configuration $\eta$ we call $j_\eta(i,j)$ the \emph{instantaneous current} across $(i,j)$.
This is obtained as the difference between the rate at which particles cross the edge from $i$ to $j$ minus the rate at which they cross the edge in the opposite direction \cite{Spohn,KL99}.
More formally  let
\begin{equation}\label{tuttoflu}
q=\sum_{i\in \Lambda_N}\sum_{s=\pm}r\left(||q^s_i||\right)q^{s}_i
\end{equation}
be the total flow of particles. This is obtained as a sum of the elementary flows associated to all the possible  elementary moves of the model.  Remember that $q$ depends on $\eta$ even if not explicitly written. Formula \eqref{tuttoflu} means that
\begin{equation}\label{tuttoflus}
q(l,m)=\sum_{i\in \Lambda_N}\sum_{s=\pm}r\left(||q^s_i||\right)q^{s}_i(l,m)\,, \qquad \forall (l,m)\in E_N\,.
\end{equation}
We then define
\begin{equation}\label{qj}
j_\eta(i,j):=q(i,j)-q(j,i)\,.
\end{equation}
The motivation and usefulness of this definition will be clear from the general argument to derive transport coefficients in section \ref{sec:DC} and it is discussed in detail in \cite{Spohn,KL99}. For any fixed configuration of particles $\eta$, the instantaneous current is antisymmetric by definition, i.e., it satisfies the symmetry $j_\eta(i,j)=-j_\eta(j,i)$.

Let $\tau_i$ denote the shift operator. It acts on configurations of particles according to $[\tau_i\eta](j)=\eta(j-i)$.
Its action on functions is then defined by $\tau_i h(\eta)=h(\tau_{-i}\eta)$. Further, let $(\Lambda_N, \mathcal E_N)$ be the undirected graph having vertices coinciding with $\Lambda_N$ and un-oriented edges $\mathcal E_N=\left\{\left\{i,j\right\}\,:\, (i,j)\in E_N\right\}$. A configuration of particles $\eta$ individuates clusters of particles as follows.  A cluster is the subgraph of $(\Lambda_N, \mathcal E_N)$ induced by the vertices belonging to the cluster. A site $i$ belongs to a cluster of particles if $\eta(i)=1$. Two sites $i,j$ belong to the same cluster if there exists a path $i^0,i^1,\dots ,i^n$ such that $\eta(i^l)=1$, $l=0,\dots,n$ and $\{i^l,i^{l+1}\}\in \mathcal E_N$, $l=0,\dots,n-1$. We denote by $C_\eta(i) \subseteq (\Lambda_N, \mathcal E_N)$ the cluster containing the particle at $i$ in the configuration $\eta$. We denote by $|C|$ the number of vertices belonging to the cluster $C$. With $\mathcal C_\eta$ we denote the collection of clusters in the configuration $\eta$ and with $|\mathcal C_\eta|$ its cardinality.

Given a configuration $\eta$ and a subset of the lattice $W$ we denote by $\eta_W$ the restriction of the configuration to the subset $W$. Given two configurations $\eta,\xi$ and two subsets $W,W'$ such that $W\cap W'=\emptyset$, we denote by $\eta_W\xi_{W'}$ the configuration of particles on $W\cup W'$ that coincides with $\eta$ on $W$ and coincides with $\xi$ on $W'$. A function $f:\{0,1\}^{\mathbb Z}\to \mathbb R$ is called local if there exists a finite subset $W\subseteq \mathbb Z$ such that $f\left(\eta_W\xi_{W^c}\right)=f(\eta)$ for any $\eta,\xi$. We denoted by $W^c$ the complementary set of $W$. The minimal subset $W$ for which this holds is called the domain of dependence of the function. This means that a function is local if the corresponding domain of dependence is finite.

A stochastic model of particle systems is called of \emph{gradient type} if there exists a local function
$h$ such that
\begin{equation}\label{gr-type}
j_\eta(i,j)=\tau_ih(\eta)-\tau_jh(\eta)\,.
\end{equation}
The relationship \eqref{gr-type} holds on the infinite lattice $\mathbb Z$ and also on the ring of size $N$ large enough with respect to the domain of dependence of the local function $h$. We will show that the k-SP is a model of gradient
type according to definition \eqref{gr-type}. In general if we consider a model with a generator of the form \eqref{gen-strong-k} but on the infinite lattice, then the corresponding instantaneous current satisfies a relationship of the form \eqref{gr-type}, but the function $h$ is not local. The locality is substituted by the following properties. The function $h$ is well defined on configurations for which the cluster of particles containing the origin is finite. For a configuration $\eta$ of this type there exists a finite subset $W$ (depending on $\eta$) such that $h(\eta_W\xi_{W^c})=h(\eta)$ for any $\xi$. Correspondingly we will have a relationship of the form \eqref{gr-type} also on a finite ring but the function $h$
now depends on the size $N$ of the lattice. When $N$ is large this sequence of functions converges to the function obtained on the infinite lattice. For this reason it is the latter that is relevant to determine the scaling limit.

We define $R(0)=0$ and
\begin{equation}
\label{Rn:def}
R(n):=\sum_{k=1}^n r(k)
\end{equation}
for $n\geq 1$; here $r(k)$ are the coefficients appearing in \eqref{gen-strong-k}. To determine the current \eqref{qj} for the model \eqref{gen-strong-k}, let us take the bond $(i,i+1)$ and compute $q(i,i+1)$ and $q(i+1,i)$. The first term is obtained by considering the flows $q_j^+$ generated by jumps to the right of particles belonging to the same cluster of $i$ and staying on the left. This yields
\begin{equation}\label{eq}
q(i,i+1)=\sum_{n=1}^{d_i^-}r\Big(d_{i+1}^++n\Big)=R\Big(d_i^-+d_{i+1}^+\Big)
-R\Big(d_{i+1}^+\Big)\,.
\end{equation}
The right-hand side of this formula is well defined and it gives the correct value (zero) also in the case of $d_i^-=0$.

Likewise considering the flows $q_j^-$ generated by jumps to the left of particles belonging to the same cluster of $i+1$ and staying on the right we get
\begin{equation}
\label{eq2}
q(i+1,i)=\sum_{n=1}^{d_{i+1}^+}r\Big(d_{i}^-+n\Big)=R\Big(d_i^-+d_{i+1}^+\Big)
-R\Big(d_{i}^-\Big)\,.
\end{equation}
Again the right-hand side of this formula is well defined and it gives the correct value (zero) also in the case of $d_{i+1}^+=0$.
The current across the edge $(i,i+1)$ is obtained as the difference between \eqref{eq} and \eqref{eq2} and it reads
\begin{equation}
\label{evai}
j_\eta(i,i+1)=R\Big(d_{i}^-\Big)-R\Big(d_{i+1}^+\Big)\,.
\end{equation}

Equation \eqref{evai} can be re-written in the form of Eq.~\eqref{gr-type}. This is not immediately evident, however. To recast \eqref{evai} into the gradient form \eqref{gr-type} we first deduce, using a direct computation, the identity
\begin{equation}
\label{curclu}
\sum_{i: \{i,i+1\}\cap C\neq \emptyset}j_\eta(i,i+1)=0
\end{equation}
that holds for any cluster of particles $C$. The identity \eqref{curclu} is essentially a consequence of the fact that both the shape of each cluster and the dynamics are invariant by an inversion of the space. Equation \eqref{curclu} says that the sum of the current on each cluster is identically zero. The fact that the model is of gradient type could be then deduced by the general arguments in \cite{DCG}. We can however in this case do a direct computation. Equation \eqref{curclu} allows to introduce a function $h_\eta(i)$ defined as follows. We fix $h_\eta(i)=0$ for any $i$ such that $\eta(i)=0$. The value inside each cluster of particles is instead fixed integrating the current on the cluster with a minus sign.
More precisely when $\eta(i)=1$ we have
\begin{equation}\label{vv}
h_\eta(i)=-\sum_{m=1}^{d_i^-}j_\eta(i-m,i-m+1)\,,
\end{equation}
Using \eqref{evai} we arrive at
\begin{equation}
\label{ggg}
h_\eta(i)=\left\{
\begin{array}{ll}
0\,, & \textrm{if} \ \eta(i)=0\,,\\
\sum_{n=0}^{d^-_i-1}\left[R\Big(\left|C_\eta(i)\right|-n\Big)-R\Big(n\Big)\right]\,, & \textrm{if} \ \eta(i)=1\,.
\end{array}
\right.
\end{equation}
Since the current is zero on bonds not intersecting any cluster and since we have \eqref{curclu} we obtain
\begin{equation}
\label{quasigrad}
j_\eta(i,i+1)=h_\eta(i)-h_\eta(i+1)\,, \qquad \forall\, i\in \Lambda_N\,.
\end{equation}
We define
\begin{equation}
\label{H:def}
H(0)=H(1)=0, \quad H(k):=\sum_{n=1}^{k-1}R(n) \quad\text{for} \quad k\geq 2.
\end{equation}
Massaging \eqref{ggg} we realize that it can be re-written as
\begin{equation}
\label{nvv}
h_\eta(i)=H\Big(d_i^-+d_i^+\Big)-H\Big(d_i^-\Big)-H\Big(d_i^+\Big)\,.
\end{equation}
The above expression is valid also when $\eta(i)=0$. Writing $h(\eta):=h_\eta(0)$ we obtain $h_\eta(i)=\tau_i h(\eta)$ and finally \eqref{quasigrad} becomes the gradient condition \eqref{gr-type} with
\begin{equation}
\label{nvvgr}
h(\eta)=H\Big(d_0^-+d_0^+\Big)-H\Big(d_0^-\Big)-H\Big(d_0^+\Big)\,.
\end{equation}
For example, for the basic SP corresponding to the generator \eqref{gen-strong}, the function $R(n)$ defined by Eq.~\eqref{Rn:def} becomes  $R(n)=n$, and therefore
\begin{equation}
\label{bella2}
h(\eta)=d^-_0(\eta)d^+_{0}(\eta)\,.
\end{equation}
This function $h$ is not local and its definition depends on the length $N$ of the lattice. Moreover it can be defined also in the infinite lattice $\mathbb Z$ but its value depends on the values assumed by the configuration $\eta$ on sites arbitrarily far from $0$. In this sense, the generalized SP process satisfies a generalized gradient condition \eqref{gr-type} with a non local function $h$ satisfying, however, the properties discussed after \eqref{gr-type}.

In the case of the $k$-SP, the function $h$ defined by \eqref{nvv} is constant on configurations for which both $d_0^\pm$ are $\geq k+2$. This means that $h$ can assume just a finite number of values and is a local function with domain of dependence the interval $[-k,k]$ containing $2k+1$ lattice sites.  This means that for the $k$-SP equation \eqref{gr-type} holds with a local function $h$. This happens for any model for which the function $r$ is different from zero just on a finite number of
natural numbers.

\section{Asymmetric models}

In this section we introduce an asymmetric version of our models switching on an external field.
For gradient stochastic lattice gases with nearest neighbors exchange dynamics it happens that the invariant measure is unchanged as far as the external field satisfies a discrete divergence free condition \cite{BDGJLstoc-int}. This is the case also for our models. In one dimension this corresponds to constant external fields.

In this case it is convenient to introduce a perturbed version of the rates as follows. Let $F$ be the constant value of the field.
The perturbed rates of jump from the configuration $\eta$ to the configuration $\sigma_i^\pm\eta$ are defined as
\begin{equation}\label{weak}
r\left(||q_i^\pm||\right)\Big(1\pm F ||q_i^\pm||\Big)\,.
\end{equation}
This means that the original rates are perturbed by a term proportional to the work done by the field in the elementary
transformation (i.e. a particle jump). The field $F$ should be small enough so that the rates in \eqref{weak} are always non negative.
When $N$ is becoming large it is then natural to consider weakly asymmetric fields of the form $F/N$ with $|F|<1$.

In presence of an external field the model will not be reversible anymore. The condition to impose assuring that a uniform measure is still invariant for the perturbed rates is that 
\begin{eqnarray}
& &\sum_{i}\sum_{s=\pm}r\left(||q_i^s(\eta)||\right)\Big[1+s F ||q_i^s(\eta)||\Big]=\nonumber \\
& &\sum_{i}\sum_{s=\pm}r\left(||q_{i-sd_i^{-s}(\eta)}^s(\sigma_i^{-s}(\eta))||\right)
\Big[1-sF||q_{i-sd_i^{-s}(\eta)}^s(\sigma_i^{-s}(\eta))||\Big]
\label{labella}
\end{eqnarray}
for any configuration $\eta$. By definition $\sigma^s_{i-sd_i^{-s}(\eta)}\left(\sigma^{-s}_i\eta\right)=\eta$, and then $||q_{i-sd_i^{-s}(\eta)}^s(\sigma_i^{-s}(\eta))||=||q_i^{-s}(\eta)||$. The right-hand side of
\eqref{labella} becomes
$$
\sum_{i}\sum_{s=\pm}r\left(||q_i^s(\eta)||\right)\Big[1-s F ||q_i^s(\eta)||\Big]\,.
$$
The validity of \eqref{labella} then follows by the left/right symmetry of the model that implies $\sum_ig\Big(||q_i^+(\eta)||\Big)=\sum_ig\Big(||q_i^-(\eta)||\Big)$ for any real function $g$.

\smallskip

We end this section computing the instantaneous current $j^F$ for an asymmetric one-dimensional  model
with a constant external field $F$. This computation will be useful to determine the mobility. Using the form of the rates \eqref{weak} after some computations similar to the ones of the previous section we obtain
\begin{equation}\label{ist-curr-E}
j^F_\eta(i,i+1) = j_\eta(i,i+1)+F\Big[2\Gamma(d_i^-+d_{i+1}^+)-\Gamma(d_i^-)-\Gamma(d_{i+1}^+)\Big]\,,
\end{equation}
where
\begin{equation}
\label{Gamma:def}
\Gamma(k):=\sum_{j=1}^kjr(j)\,.
\end{equation}

\section{Transport coefficients}
\label{sec:DC}

For gradient and reversible models the transport coefficients can be computed explicitly using just the static part of Green-Kubo formulas \cite{Spohn,KL99}. We sketch an argument showing how the instantaneous current plays a crucial role. We embed the lattice $\Lambda_N$ into the continuous one-dimensional circle $[0,1]$ with periodic boundary conditions. This is done by associating the point $i/N\in[0,1]$ to the lattice site $i\in \Lambda_N$. Let us consider a weakly asymmetric SP process. This means that the system of particles is subject to a constant external field $F/N$, i.e. the strength of the field depends on the parameter $N$ and in particular is $O(1/N)$. The natural scaling limit for this class of processes is the diffusive one. For this reason we consider a model with generator
$$
\sum_{i\in \Lambda_N}\sum_{s=\pm}r\left(||q_i^s||\right)\Big(1+s \frac FN ||q_i^s||\Big)\big[f\big(\sigma^{s}_{i}\eta\big)-f\big(\eta\big)\big]
$$
and then we multiply by a factor of $N^2$ the rates. Let $\eta_t$ be the random configuration of particles at time $t$. Let also $\mathcal J_t(i,j)$ be the total current flown across the bond $(i,j)$ in the time window $[0,t]$. This is the total number of particles that have flown from $i$ to $j$ minus the total number of particles that have flown from $j$ to $i$. A simple argument \cite{Spohn} shows that
\begin{equation}
\label{mart}
\mathcal J_t(i,j)-N^2\int_0^tj_{\eta_s}^{F/N}(i,j)\,ds\,,
\end{equation}
is a martingale that can be neglected in the scaling limit. The factor $N^2$ is due to the fact that we are speeding up the process.

Let $W:[0,1]\to \mathbb R$ periodic be a smooth vector field on the continuous circle.
The total work done in this time window by the vector field $W$ is proportional to
\begin{equation}\label{total-work}
\frac {1}{2N}\sum_{(i,j)\in E_N}\mathcal J_t(i,j)\mathcal W(i,j)\,,
\end{equation}
where
\begin{equation}
\label{disc-vec}
\mathcal W(i,i+1)=\frac 1N\int_0^1d\,z\, W\left(\frac{i+z}{N}\right)\,.
\end{equation}
Note that $\mathcal W$ is a discrete vector field defined on the bonds of the graph. More precisely, $\mathcal W(i,j)$ is defined only if $\{i,j\}\in \mathcal E_N$ and it satisfies $\mathcal W(i,j)=-\mathcal W(j,i)$. This corresponds to the fact that if we orient as $(i,j)$ the un-oriented bond $\{i,j\}$  we have the value $\mathcal W(i,j)$ while we would have the opposite value if we orient the bond in the opposite direction. We emphasize that $\mathcal W$ is the discretize version of the continuous vector field $W$.
Using \eqref{mart} we can write \eqref{total-work} up to a neglecting martingale term as
\begin{equation}
\label{work-up}
\frac{N}{2}\int_0^t\sum_{(i,j)\in E_N}j^{F/N}_{\eta_s}(i,j)\mathcal W(i,j)ds\,.
\end{equation}
Recalling \eqref{ist-curr-E} and introducing the function
\begin{equation}
\label{func-g}
g(\eta)=2\Gamma(d_0^-+d_{1}^+)-\Gamma(d_0^-)-\Gamma(d_{1}^+)
\end{equation}
we can write \eqref{work-up}  as
\begin{equation}
N\int_0^t\sum_{i}j_{\eta_s}(i,i+1)\mathcal W(i,i+1)ds+F\int_0^t\sum_{i}\tau_ig(\eta_s)\mathcal W(i,i+1)ds\,.
\label{primo-passo}
\end{equation}
In contrast to \eqref{work-up}, a factor  $1/2$ factor is missing in \eqref{primo-passo}, and correspondingly we are not summing over edges of the type $(i,i-1)$; by antisymmetry the result is the same.
Since the unperturbed current $j_\eta$ is gradient formula \eqref{primo-passo} becomes after a discrete integration by parts
\begin{equation}
N\int_0^t\sum_{i}\tau_i h(\eta_s)\nabla\cdot \mathcal W(i)ds+ F \int_0^t\sum_{i}\tau_ig(\eta_s)\mathcal W(i,i+1)ds\,.
\label{secondo-passo}
\end{equation}
In the above equation we denote by
$$
\nabla\cdot \mathcal W(i):=	\sum_{j:(i,j)\in E_N}\mathcal W(i,j)
$$
the discrete divergence of the discrete vector field $\mathcal W$. We use the same symbol $\nabla \cdot$ both to denote the discrete and the continuous divergence since they are closely related objects (and also strictly related to the divergence of a flow that share also the same notation). There is no risk of confusion since every time $\nabla \cdot $ is applied to a discrete vector field has to be understood as a discrete divergence while instead has to be interpreted as a continuous divergence when it is applied to a continuous vector field. Since the vector field $W$ is smooth, recalling its definition \eqref{disc-vec}, we obtain $N\nabla\cdot \mathcal W(i)=\frac 1N\nabla\cdot W(i/N)$ up to higher order terms. We apply also the
replacement Lemma that roughly says
\begin{equation}
\label{caldo}
\frac 1N \sum_i\int_0^t\tau_if(\eta_s)\psi\left(\frac iN\right)ds\simeq \int_{[0,1]}dx\,\int_0^t\mathbb E_{\mu^{\lambda[\rho(x,s)]}}[f]\psi(x)ds\,,
\end{equation}
for any local function $f$ and smooth function $\psi:[0,1]\to \mathbb R$. In this formula $\mu^{\lambda[\rho]}$
is the grand-canonical invariant measure of the dynamics with chemical potential fixed by the local density of particles $\rho$.
The symbol $\simeq$ means that in the limit $N\to +\infty$ we can substitute the left-hand side of \eqref{caldo} by the right-hand side and this
allows us to get a closed equation for the density field in the limit. This is proved rigourously for models of the type discussed here
in \cite{sing_diff}.

For the SP, we have $\mu^{\lambda[\rho]}=\nu_\rho$ where $\nu_\rho$ is the Bernoulli product measure with parameter $\rho$. We define
\begin{equation}
\label{finalmente}
G(\rho)=\mathbb E_{\nu_{\rho}}\left(h(\eta)\right)\quad\text{and} \quad \sigma(\rho)=\mathbb E_{\nu_{\rho}}\left[g(\eta)\right].
\end{equation}
With high probability in the $N\to \infty$ limit, \eqref{secondo-passo} converges to
\begin{equation}\label{cenone}
\int_0^t ds\int_{[0,1]}dx\,\left[G(\rho(x,s))\nabla\cdot W(x)+F\sigma(\rho(x,s))W(x)\right]\,,
\end{equation}
for any test vector field $W$.
Formula \eqref{cenone} says, in a weak form, that the typical limiting current is
\begin{equation}
\label{cenonef}
J(\rho):=-D(\rho)\nabla\rho+\sigma(\rho)F
\end{equation}
where the diffusion coefficient is
\begin{equation}\label{diff}
D(\rho)= \frac{ d G(\rho)}{d \rho}\,,
\end{equation}
while the mobility $\sigma$ is given in \eqref{finalmente}. The typical current \eqref{cenonef}
together with the continuity equation give the hydrodynamic behavior \eqref{DEE}.

\smallskip
To compute the transport coefficients for SP we recall again that in this case $\mu^{\lambda[\rho]}=\nu_\rho$ is a product Bernoulli measure with parameter $\rho$. Using this fact we establish
that $d_0^-$ and $d_1^+$ are independent geometric random variables with parameter
$\rho$, that is $\mathbb P\left(d_i^\pm=k\right)=(1-\rho)\rho^k$, $k=0,1,2,\dots$. Given a sequence $a(n)$ we denote by
$$
S_a(\rho):=\sum_{j=0}^{+\infty}\rho^ja(j)
$$
the corresponding power series and use a simple notation $S(\rho)=S_r(\rho)$ for the power series corresponding to the sequence $r(n)$.

First, we compute the mobility. We have
$$
\mathbb E_{\nu_\rho}\left(\Gamma(d_0^-)\right)=\mathbb E_{\nu_\rho}\left(\Gamma(d_1^+)\right)=(1-\rho)S_\Gamma(\rho)\,.
$$
We then notice that
\begin{eqnarray}
& &\mathbb E_{\nu_\rho}\left(\Gamma(d_0^-+d_1^+)\right)=(1-\rho)^2\sum_{k=0}^{+\infty}
\sum_{l=0}^{+\infty}\rho^{k+l}\Gamma(k+l)\nonumber \\
& &=(1-\rho)^2\sum_{j=0}^{+\infty}\rho^j(j+1)\Gamma(j)=(1-\rho)^2\frac{d}{d\rho}\left(\rho S_\Gamma(\rho)\right)\,.
\end{eqnarray}
Recalling the definition \eqref{Gamma:def} of $\Gamma$ we obtain $S_\Gamma(\rho)=\frac{\rho}{1-\rho}\frac{d S}{d \rho}(\rho)$. Putting everything together we arrive at
\begin{equation}
\label{capelli}
\sigma(\rho)=2\rho(1-\rho)\frac{d}{d \rho}\left(\rho \frac{d S}{d\rho}(\rho)\right)\,.
\end{equation}

\begin{figure}
\begin{center}
 \includegraphics[width=99mm,height=80mm]{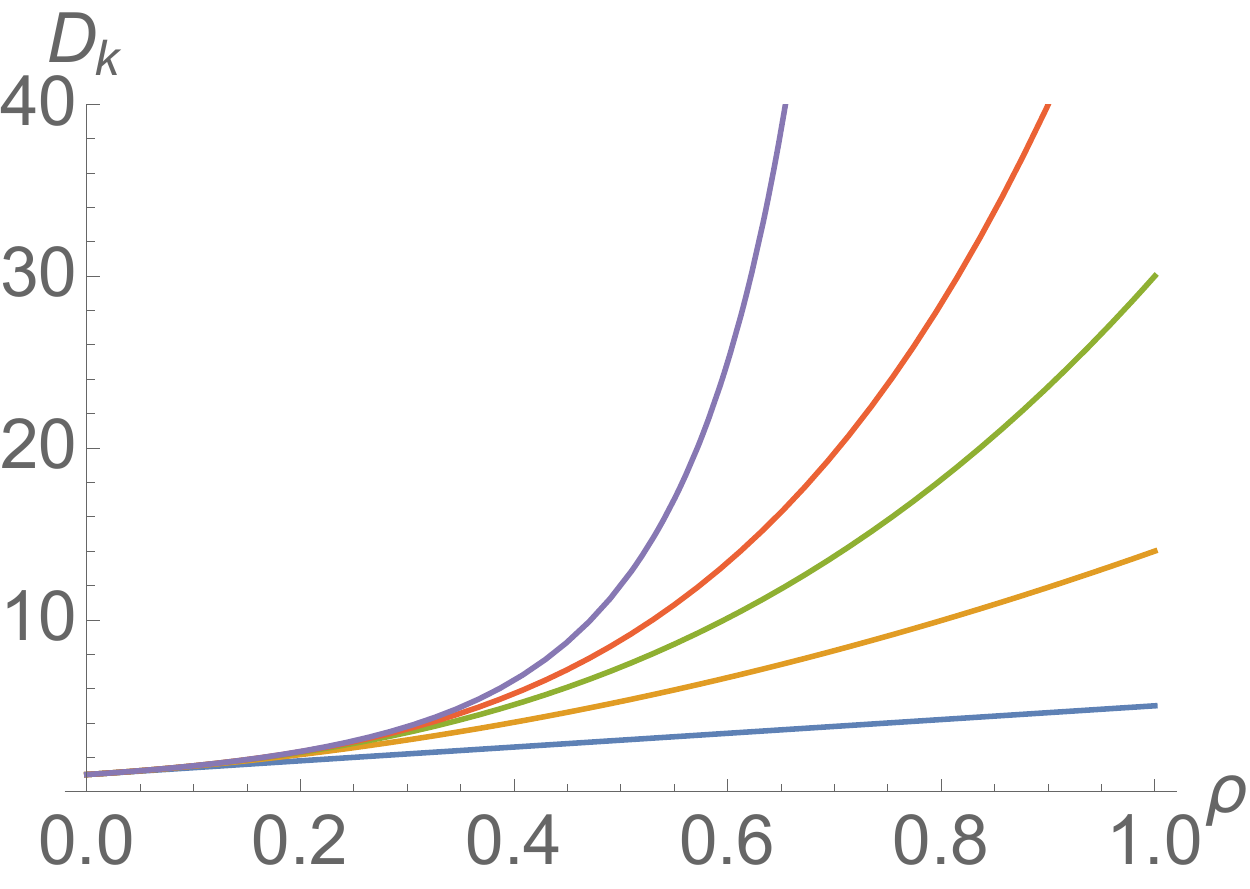}
 \caption{The diffusion coefficient $D_k$ of the k-SP given by \eqref{fine} for $k=1,2,3,4,\infty$ (from bottom to top).}
\label{fig:Dk}
\end{center}
\end{figure}

\smallskip
We compute now the diffusion coefficient. We use the identity
$$H(d_0^-+d_0^+)=\chi(d_0^-\neq 0)H(d_0^-+d_1^++1)$$
for $H$ defined by \eqref{H:def}. Recalling that $d_0^-$ and $d_1^+$ are independent geometric random variables of parameter $\rho$ we obtain
\begin{eqnarray}
\mathbb E_{\nu_\rho}\left(H(d_0^-+d_0^+)\right)
&=&(1-\rho)^2\sum_{l=1}^{+\infty}\sum_{j=0}^{+\infty}\rho^{l+j}H(l+j+1) \nonumber\\
&=&(1-\rho)^2\sum_{k=1}^{+\infty}\rho^kkH(k+1)  \nonumber\\
&=&(1-\rho)^2\left(\frac{d S_H}{d \rho}(\rho)-\frac{S_H(\rho)}{\rho}\right)\,.
\end{eqnarray}
Using this computation and adding the averages of the other terms we get
$$
G(\rho)=(1-\rho)^2\left(\frac{d S_H}{d \rho}(\rho)-\frac{S_H(\rho)}{\rho}\right)-2(1-\rho)S_H(\rho)\,.
$$
After straightforward calculations we find $S_H(\rho)=\frac{\rho S(\rho)}{(1-\rho)^2}$ from which we deduce
a neat formula
$$
G(\rho)=\rho \frac{d S}{d \rho}(\rho)
$$
that gives finally
\begin{equation}\label{kap}
D(\rho)=\frac{d}{d \rho}\left(\rho \frac{d S}{d \rho}(\rho)\right)=\sum_{k=1}^\infty k^2\rho^{k-1}r(k)\,.
\end{equation}
This coincides with the formula in \cite{sing_diff}. Our derivation elucidates the non-trivial discrete gradient structure behind.
For the basic SP with $r(k)\equiv 1$ we get
\begin{equation}
\label{diffSP}
D(\rho)=\frac{1+\rho}{(1-\rho)^3}\,.
\end{equation}
For the k-SP with rates given by \eqref{strong-k}
we have
\begin{equation}
\label{fine}
D_k(\rho) = \sum_{n=0}^k (n+1)^2\rho^n\,,
\end{equation}
that is the $k^{\text{th}}$ order Taylor polynomial around zero of \eqref{diffSP}. The diffusion coefficient $D_k(\rho)$ is an increasing function of both $\rho$ and $k$ (see Fig.~\ref{fig:Dk}). For $k=0$ the diffusion coefficient is constant, $D_0=1$, since the 0-SP model reduces to the SEP.

Recalling that in the case of Bernoulli product measures the free energy is given by \eqref{fff}
we can check directly the validity of the Einstein relation \eqref{FDT}  between the transport coefficients. In particular the corresponding mobilities are obtained after multiplying \eqref{diffSP} and \eqref{fine} by $2\rho(1-\rho)$.

\section{Related models}

In this section we discuss some models related to the SP and show how to deduce their transport coefficients starting from the established transport coefficients for the SP. For concreteness, we consider the basic SP process; the general case could be discussed very similarly.

\subsection{Leap frog model}\label{LFMP} In this model every particle jumps with rate one to the first empty site on its right and to the first empty site on its left. This means that isolated particles perform a simple random walk but particles belonging to clusters jump with rate one to the two boundary sites of the cluster. If we forget the labels of the particles and observe just the evolution of the unlabeled particles the dynamics is exactly the same as for the basic SP. This means that the leap frog model and the SP are two different dynamics on labelled particles giving as a result the same stochastic evolution for indistinguishable particles. Clearly the bulk transport coefficients are the same as in the SP. These models can be considered likewise in the case of a rate $r(k)$ for a jump of $k$ steps and again they give the same unlabeled dynamics as the SP with the same rates. This is the dynamics discussed in \cite{sing_diff}.

\subsection{ Hammersley process} A natural change of perspective on a stochastic lattice gas is obtained by interpreting particles as empty sites and empty sites as particles. This is called the particle-hole symmetry. In the case of the SP it is not difficult to see that the stochastic model obtained in this way is nothing else that the Hammersley process (see for example \cite{Pab}). In this model particles cannot overtake but are allowed to perform jumps of arbitrary length. In particular every particle jumps with unitary rate to every empty site that can reach without going over other particles. Again a generalization to rates $r(k)$ is possible. A configuration $\eta_H$ of the  Hammersley process is obtained as
$\eta_H=1-\eta$ where $\eta$ is a configuration of SP. When $\eta$ evolves according to SP dynamics then $\eta_H$ evolves according to the Hammersley process. We can then easily obtain the transport coefficients for the Hammesrley process from the ones for SP. Since macroscopically the densities will again satisfy  $\rho_H+\rho=1$ we can deduce
$$
\partial_t\rho_H=-\partial_t \rho=-\nabla\cdot D(\rho)\nabla \rho=\nabla\cdot D(1-\rho_H)\nabla \rho_H\,.
$$
Since we write the macroscopic equation for the Hammersley process as
$$
\partial_t\rho_H=\nabla\cdot D_H(\rho_H)\nabla \rho_H\,,
$$
we obtain, using \eqref{diffSP}, the diffusion coefficient for the  Hammersley process as
\begin{equation}
\label{D:HP}
D_H(\rho_H)=D(1-\rho_H)=\frac{2-\rho_H}{\rho_H^3}\,.
\end{equation}

\subsection{Exclusion with avalanches}\label{ewa}

In this subsection we will use an argument that needs boundary conditions different from the periodic ones that we use otherwise throughout all the paper. Here we look at an exclusion process with nearest-neighbor hopping and avalanches \cite{EPA}. Informally this dynamics can be described as follows. Particles satisfy an exclusion principle and when isolated perform simple random walks. The particles are however elastic so that if a particle joins a cluster, the particle on the other extremum of the cluster is immediately pushed one step away. If this second particle joins another cluster the same happens for this second cluster and so on. This transformation is instantaneous and can generate avalanches. On a one-dimensional ring when the number of particles is less or equal to the numbers of holes the dynamics is well defined. Whatever is the initial configuration, after a transient regime there will be just isolated particles evolving. We consider just the dynamics after the transient regime. In one dimension it is possible to map locally the basic SP to the exclusion with avalanches. The mapping is the following
\begin{equation}
\begin{array}{ccc}
\bullet & \Longleftrightarrow &   \sqsupset \,\stackrel{}{\blacksquare}\, \sqsubset  \\
\circ & \Longleftrightarrow & \sqsupset\,\sqsubset
\end{array}
\label{mapusauro}
\end{equation}
In the above picture $\bullet$ and $\circ$ are respectively particles and empty site for the SP process, while
$\blacksquare$ and $\square$ are respectively particles and empty sites for exclusion with avalanches. With $\sqsubset$ and $\sqsupset$ we represent half empty sites. The configuration of particles for exclusion with avalanches is built up
following the rule
$$\sqsubset\sqsupset \ \ \ \Longleftrightarrow\ \ \  \square$$
that means that two half empty sites glue together to form an entire empty site.
This is an example of a configuration of the SP and the corresponding configuration of particles for the exclusion process with avalanches
$$
\bullet\bullet\circ\bullet\circ\circ\bullet\qquad \Longleftrightarrow\qquad  \sqsupset\,\stackrel{}{\blacksquare}\,\stackrel{}{\square}\,\stackrel{}{\blacksquare}\,\stackrel{}{\square}\,\stackrel{}{\square}\,
\stackrel{}{\blacksquare}\,\stackrel{}{\square}\,\stackrel{}{\square}\,\stackrel{}{\square}\,\stackrel{}{\blacksquare}\, \sqsubset
$$
When a configuration of round particles evolves according to the SP dynamics, the corresponding configuration of squared particles evolves according to the exclusion process with avalanches. We stress that when we speak of configurations of the exclusion with avalanches we are referring to the configurations with isolated particles obtained after the transient period.

We underline the fact that on the ring this map is a local map and not a global one since for example on the ring it is not possible to have a one to one correspondence between particles following the SP dynamics and particles evolving according to the exclusion process with avalanches. This is because the number of sites of the two lattices are different. Indeed, if we consider a ring with $N$ sites where $M$ strong particles are evolving, in the corresponding exclusion process with avalanches $M$ particles are evolving on a ring with $N+M$ sites. In the latter process, only the configurations with isolated particles are possible but still the numbers of possible configurations do not match. (For example, when $N=2$ and $M=1$, there are two possible configurations for the SP and three configurations for the exclusion process with avalanches.)

This local mapping can be transformed into a global one on a interval with closed boundaries. Since we are interested in transport coefficients this is enough for us. Consider a one-dimensional lattice with $N$ sites with $M<N$ strong particles evolving. At the boundary sites $1,N$ the dynamics is blocked: A particle at $1$ cannot jump to the left and a particle at $N$ cannot jump to the right. Consider now any configuration of SP in this interval and apply the map \eqref{mapusauro}. If we add a $\sqsubset$ on the left and a $\sqsupset$ on the right of a configuration, we get a configuration of exclusion with avalanches with $M$ particles on a lattice with $N+M+1$ sites. The mapping is now globally one to one and the dynamics of SP with zero flux outside the boundary is mapped into the exclusion with avalanches with reflecting boundary. Indeed for the exclusion with avalanches model the sites labeled $1$ and $N+M+1$ are always empty. This is because the boundary are reflecting and a particle that reach one of these two sites bounces immediately back.

We consider the limiting behavior when $M=\lfloor \alpha N\rfloor$ and $N$ diverge, and the lattice of the SP dynamics is embedded on the unit interval $[0,1]$. In this case the empirical measure for the SP process will follow in the limit a nonlinear diffusion with diffusion coefficient \eqref{diffSP}. Recalling that the current will converge to \eqref{cenonef} with $F=0$ and that the flux of particles across the boundary is zero, we deduce that the boundary conditions to be added to the equation are $\nabla\rho(0,t)=\nabla\rho(1,t)=0$. Correspondingly the empirical measure associated to the exclusion with avalanches will converge to a density $\rho_A$ that satisfies a nonlinear diffusion equation with diffusion coefficient $D_A$. Also for this model there is no flux of particles across the boundaries so that $\nabla \rho_A=0$ at the boundary. Since the mesh of the lattice is $\frac 1N$ and the lattice has $N+\lfloor \alpha N\rfloor+1$ sites, the macroscopic region where the limiting evolution of the exclusion with avalanches model is defined is the interval $[0,1+\alpha]$. The boundary conditions are then $\nabla \rho_A(0,t)=\nabla\rho_A(1+\alpha,t)=0$. Our goal is to determine $D_A$ from \eqref{diffSP} and the mapping \eqref{mapusauro} recovering in this way the result of \cite{EPA}. Let $\eta$ be the configuration of the SP dynamics and $\eta_A$ the corresponding configuration for the exclusion with avalanches obtained by the mapping \eqref{mapusauro}. Using the definition we obtain
\begin{equation}
\label{gel}
\sum_{j=1}^i\eta(j)=\sum_{j=1}^{i+\sum_{k=1}^i\eta(k)}\eta_A(j)\,.
\end{equation}
Formula \eqref{gel} can be thought as the definition of the mapping between the two models and it is equivalent
to the mapping \eqref{mapusauro}. For $y\leq z$, we have
$$
\frac 1N \sum_{i\in [yN,zN]}\eta(i)\stackrel{N\to\infty}{\to} \int_y^z\rho(x)dx=F(z)-F(y)
$$
where $F(y):=\int_0^y\rho(x)dx$ is the distribution function associated to the density $\rho$. Likewise for the exclusion process with avalanches model 
$$
\frac 1N \sum_{i\in [yN,zN]}\eta_A(i)\stackrel{N\to+\infty}{\to} \int_y^z\rho_A(x)dx=F_A(z)-F_A(y)
$$
where $F_A(y):=\int_0^y\rho_A(x)dx$ is the distribution function associated to the density $\rho_A$.
If we multiply both sides of \eqref{gel} by $\frac 1N$ and consider $i=\lfloor xN\rfloor$, we obtain a relationship between the empirical measures that in the limit becomes
\begin{equation}
\label{gelcont}
F(x)=F_A\left(x+F(x)\right)\,.
\end{equation}
Equation \eqref{gelcont} holds for any pairs $F$ and $F_A$ obtained by scaling limits of empirical measures of particles configurations $\eta$ and $\eta_A$ related by \eqref{mapusauro} or equivalently by \eqref{gel}.

We now write $F=F(x,t)=\int_0^x\rho(y,t)dy$ where $\rho(x,t)$ is the solution of the hydrodynamic equation for SP on the interval $[0,1]$ and having Neumann boundary conditions. We also define $F_A=F_A(x,t)$ via Eq.~\eqref{gelcont} with $F(x)$ substituted by $F(x,t)$. By construction, $F_A(x,t)=\int_0^x\rho_A(y,t)dy$ where $\rho_A(x,t)$ is the solution of the hydrodynamic equation for exclusion with avalanches on the interval $[0,1+\alpha]$. Since for any time $t$ the relation
\begin{equation}
\label{gelcontt}
F(x,t)=F_A\left(x+F(x,t),t\right)\,,
\end{equation}
is satisfied we can differentiate it with respect to time to yield
\begin{equation}\label{boov}
\partial_t F=\partial_t F_A+\rho_A\partial_t F\,.
\end{equation}
To simplify notation we are not writing the arguments of the functions. Every time that we have a function with index $A$ the argument is $(x+F(x,t),t)$ while every time that there is a function without the lower index $A$ the argument is simply $(x,t)$.
Differentiating \eqref{gelcontt} with respect to $x$  we get
\begin{equation}
\label{gu}
\rho=\rho_A(1+\rho)\,.
\end{equation}
Differentiating \eqref{gu} with respect to $x$ and keeping in mind that the arguments of functions $\rho$ and $\rho_A$ are different, we deduce
\begin{equation}
\label{f1}
\nabla\rho=\frac{\nabla\rho_A}{(1-\rho_A)^3}\,.
\end{equation}
Using the boundary conditions and the nonlinear equations satisfied by the densities we get
\begin{equation}\label{al}
\partial_t F=D(\rho)\nabla \rho \qquad \textrm{and} \qquad \partial_tF_A= D_A(\rho_A)\nabla \rho_A\,.
\end{equation}
Using all together \eqref{boov}, \eqref{gu}, \eqref{f1} and \eqref{al} we obtain
\begin{equation}\label{ucci}
D_A(\rho_A)=D\left(\frac{\rho_A}{1-\rho_A}\right)\frac{1}{(1-\rho_A)^2}\,.
\end{equation}

This relationship is true for any SP dynamics with any function $r$. It is possible then to obtain the diffusion coefficient
for an exclusion with avalanches model on which avalanches of size $k$ (i.e. avalanches on which exactly $k$ particles are moving) happens with a rate $r(k)$. For example, take the 0-SP, i.e., the simple exclusion process. In the corresponding exclusion process with avalanches the only interaction is given by the constraint that two particles have to be at distance bigger than $1$. Since $D=1$ for the simple exclusion process, we obtain
\begin{equation}
\label{D:RP}
D_A(\rho_A)=\frac{1}{(1-\rho_A)^2}\,.
\end{equation}
There are no avalanches in this particular exclusion process. This process is actually a repulsion process which was studied in \cite{RP13} where \eqref{D:RP} was derived via a different method.

Considering the basic SP and using the explicit expression \eqref{diffSP},  we have
\begin{equation}\label{finlay}
D_A(\rho_A)=\frac{1}{(1-2\rho_A)^3}\,,
\end{equation}
that is the result in \cite{EPA}.

\section{Strong particles with Gibbsian invariant measure}

As soon as we consider invariant measures not of product type the computational difficulties increase and the class of solvable models reduces. We will consider some special cases. Instead of studying the general case we analyze here a specific example. Namely, we consider Gibbsian measures $\mu^\lambda_N (\eta)=\frac{1}{Z}e^{-H_\lambda(\eta)}$ where the Hamiltonian is given by
\begin{equation}
\label{ham-2corpi}
H_\lambda(\eta)=\gamma\sum_{i\in \mathbb Z_N}\mathbb I\left[\eta(i)\neq \eta(i+1)\right]-\lambda\sum_{i\in \mathbb Z_N}\eta(i)\,.
\end{equation}
Here $\mathbb{I}$ is the characteristic function and we explicitly write only the dependence on the chemical potential $\lambda$ since it is the parameter that fixes the value of the density $\rho$ while $\gamma$ is a given parameter. The interpretation of the energy is simple, it corresponds to $\lambda$ times the negative number of particles plus $2\gamma$ times the number of clusters of particles $|\mathcal C_\eta|$.

We consider a process of strong particles so that when a particle tries to jump it always succeeds. We classify the elementary moves into 3 classes. Let $\eta$ be the configuration before the elementary transformation and $\eta'$ the one after.

\smallskip

1) The first class is composed by the elementary transformations for which $H_\lambda(\eta)=H_\lambda(\eta')$. These are transformations on which one single cluster is moving to the left or to the right or one cluster breaks into two pieces but one of them glues to another cluster so that the total number of clusters does not change. The rate of jump of an elementary move belonging to the first class and such that the total elementary flow associated is $n$ (this means there are $n$ particles moving of one step) is set equal to $\frac qn$, where $q$ is a fixed positive parameter.

\smallskip

2) The second class is composed by the elementary transformations for which $H_\lambda(\eta)=H_\lambda(\eta')-2\gamma$.
This happens when one cluster breaks into two clusters so that the total number of clusters increases by one.
The rate of jump of an elementary move belonging to the second class and such that the total elementary flow associated is $n$ is set equal to $\frac qn$.

\smallskip

3) The third class is composed by the elementary transformations for which $H_\lambda(\eta)=H_\lambda(\eta')+2\gamma$.
This happens when one cluster moves and glues to another cluster so that the number of clusters decreases by one. The rate of jump of an elementary move belonging to the third class and such that the total elementary flow associated is $n$ is set equal to $\frac{qe^{2\gamma}}{n}$.

There are no other possibilities. By a direct verification of the detailed balance condition one finds that the dynamics just defined is reversible with respect to a Gibbs measure with energy $H_\lambda$ for any value of the chemical potential $\lambda$.

The derivation of the transport coefficients for the above class of models is more involved since condition \eqref{curclu} is not satisfied and we cannot apply directly the construction done for the previous models. To generalize \eqref{curclu} we begin with a definition. Given a lattice site $i$ such that $\eta(i)=0$ we denote by $C_\eta(i)$ the cluster of empty sites containing the site $i$, and we call such an empty cluster a \emph{lake} if $|C_\eta(i)|=1$ and a \emph{sea} if $|C_\eta(i)|>1$. Consider for example a cluster of particles $C$ that has a lake to its left and a sea to its right. In this situation
\begin{equation}
\label{cuclu2}
\sum_{i: \{i,i+1\}\cap C\neq \emptyset}j_\eta(i,i+1)=q\left(1-e^{2\gamma}\right)\,.
\end{equation}
This result is obtained since the currents generated by the jumps of the particles that are in the interior part of the cluster
compensate exactly. The currents generated by the jumps of the extreme left and extreme right particles of the cluster are different and their difference is exactly the right hand side of \eqref{cuclu2}. It is important to observe that \eqref{cuclu2} does not depend on the size of the cluster.

Equation \eqref{cuclu2} corresponds to one of the four possible arrangements. If a cluster has a sea to its left and a lake to its right, we obtain \eqref{cuclu2} with a minus sign on the right-hand side. For a cluster that has a lake both to the right or to the left, the right-hand side of \eqref{cuclu2} should be replaced by $0$ since the currents compensate exactly, i.e., we recover \eqref{curclu}. The same happens for a cluster having a sea both on the right and on the left. Condition \eqref{curclu} is generalized in this case as follows. The basic idea is that instead of cutting particles into clusters using any empty site and setting equal to zero the value of the function $h$ on each empty site, we instead cut the lattice into subregions but using only the seas and set equal to zero the value of the function $h$ only on lattice sites belonging to seas. Let $A$ be any interval of the lattice contained inside two consecutive seas. Then 
\begin{equation}
\label{cuclu3}
\sum_{i: \{i,i+1\}\cap A\neq \emptyset}j_\eta(i,i+1)=0\,.
\end{equation}
This happens because the currents of all the clusters having on the right and on the left a lake compensate exactly
and the currents of the leftmost and rightmost clusters are the same but with opposite signs.
Proceeding as in the previous cases we can define a function $h$ as follows. If $\eta(0)=0$ and $|C_\eta(0)|>2$ this means that the origin belongs to a sea of empty sites then we fix $h(\eta)=0$. Otherwise let $m=m(\eta)$ be the first site belonging to an empty sea on the left of the origin. We then define
\begin{equation}\label{sabatotesi}
h(\eta)=-\sum_{i=m(\eta)}^{-1}j_\eta(i,i+1)\,.
\end{equation}
Using the same arguments used before we have that $j_\eta(i,i+1)=\tau_ih(\eta)-\tau_{i+1}h(\eta)$ with a function $h$
that is not local but satisfies the properties illustrated after \eqref{gr-type}.

The diffusion coefficient $D(\rho)$ can be extracted from relation $D(\rho)=G'(\rho)$ where $G(\rho)=\mathbb E_{\mu_{\lambda[\rho]}}\left(h\right)$. The expected value is with respect to the Gibbs measure on the infinite lattice $\mathbb Z$ corresponding to the two body interaction $\gamma$ and to the chemical potential $\lambda[\rho]$ that fixes the density to the value $\rho$. It is convenient to describe this measure as a stationary Markov measure on $\{0,1\}^{\mathbb Z}$ with transition matrix
$$
T=
\begin{pmatrix}
  1-\alpha & \alpha \\
  \beta & 1-\beta \\
\end{pmatrix}
$$
where the coefficients $\alpha,\beta$ are related to the interaction parameter $\gamma$ and the chemical potential $\lambda$. To see which is the relationship between  $(\gamma,\lambda)$ and $(\alpha,\beta)$ let us come back
to the ring with $N$ sites. Given a configuration $\eta$ we call
$$
\mathcal N_{\epsilon,\epsilon'}(\eta):=\sum_{i\in \mathbb Z_N}\mathbb I\left[\eta(i)=\epsilon\,,\,\eta(i+1)=\epsilon'\right]\,, \qquad \epsilon,\epsilon'=0,1\,.
$$
Then $\mathcal N_{0,1}(\eta)=\mathcal N_{1,0}(\eta)=|\mathcal C_\eta|$ where $|\mathcal C_\eta|$ is the numbers of clusters. We also have $\sum_{\epsilon,\epsilon'}\mathcal N_{\epsilon,\epsilon'}(\eta)=N$ and $\mathcal N_{0,1}(\eta)+\mathcal N_{1,1}(\eta)=\sum_{x}\eta(x)$.
Using these relationships we can write the energy \eqref{ham-2corpi} as
\begin{equation}\label{ham-n}
H_\lambda(\eta)=\left(2\gamma-\lambda\right)|\mathcal C_\eta|-\lambda \mathcal N_{1,1}\,.
\end{equation}
By a direct computation we find
\begin{equation}
\label{ilpro}
\log\prod_{i=1}^NT_{\eta(i),\eta(i+1)}=|\mathcal C_\eta|\,\log \frac{\alpha\beta}{(1-\alpha)^2}+\mathcal N_{1,1}(\eta)\log\frac{1-\beta}{1-\alpha}+N\log(1-\alpha)\,.
\end{equation}
Fixing the parameters $(\alpha,\beta)$ through the conditions
\begin{equation}
\left\{
\begin{array}{l}
\lambda=\log\frac{1-\beta}{1-\alpha} \\
\gamma=\frac 12 \log\frac{(1-\alpha)(1-\beta)}{\alpha\beta}
\end{array}
\right.
\end{equation}
we notice that apart irrelevant constant terms \eqref{ilpro} gives the same weight to any configuration as the
Hamiltonian $H_\lambda$. This means that in the limit of large N the measure $\mu_N^\lambda$ converges to
a stationary Markov measure on $\{0,1\}^\mathbb Z$ with transition matrix $T$. We used periodic boundary
conditions but the limiting behavior is independent on the boundary conditions. The invariant measure for the Markov chain is given by $\left(\frac{\beta}{\alpha+\beta},\frac{\alpha}{\alpha+\beta}\right)$. Therefore the measures $\mu^{\lambda[\rho]}$ that are used in computing the diffusion coefficient are stationary Markov measures on $\{0,1\}^\mathbb Z$ with transition matrix $T$ whose parameters are determined by
\begin{equation}\label{aep}
\left\{
\begin{array}{l}
\rho=\frac{\alpha}{\alpha+\beta} \\
\gamma=\frac 12 \log\frac{(1-\alpha)(1-\beta)}{\alpha\beta}
\end{array}
\right.
\end{equation}
The second equation is there since the parameter $\gamma$ is fixed in our model and the first equation indirectly fixes the chemical potential in such a way that the typical density is $\rho$.
The diffusion coefficient is then $D(\rho)=G'(\rho)$ where
$$
G(\rho)=\mathbb E_{\mu^{\lambda[\rho]}}\left(h\right)\,,
$$
and the function $h$ is determined by \eqref{sabatotesi}.

\smallskip

In general this computation is long and hence we discuss only a simplified case for a zero strong class of models. For these models the gradient condition is automatically satisfied if we impose that the instantaneous current is the same with opposite sign at the boundary edges of each cluster. In this case \eqref{curclu} is indeed obeyed and we can obtain the current as the gradient of a function whose value is set equal to zero on empty sites.
For zero strong particles this can be realized for example defining a function $h$ as
\begin{equation}
\label{dubai}
h(\eta)=\left\{
\begin{array}{ll}
0 & \textrm{if}\ \eta(0)=0\,, \\
a\geq 0 & \textrm{if}\  |C_\eta(0)|=1\,,\\
1 & \textrm{if}\ |C_\eta(0)|>1\,.
\end{array}
\right.
\end{equation}
The function $h$ determines the instantaneous current and since the particles are zero strong this fixes also the
rates of the jumps. To have reversibility with respect to the Gibbs measure with energy \eqref{ham-2corpi}, the constant $a$ has to be related to $\gamma$ by $a=e^{2\gamma}$. Equations \eqref{aep} become
\begin{equation}
\label{dainvertire}
\left\{
\begin{array}{l}
\rho=\frac{\alpha}{\alpha+\beta} \\
a=\frac{(1-\alpha)(1-\beta)}{\alpha\beta}
\end{array}
\right.
\end{equation}
Finally, using the fact that $\mu^{\lambda[\rho]}$ is a Markov measure, we get
\begin{eqnarray}
G(\rho)&=&\mathbb E_{\mu^{\lambda[\rho]}}(h)= a\mathbb P_{\mu^{\lambda[\rho]}}\big(|C_\eta(0)|=1\big)+\mathbb P_{\mu^{\lambda[\rho]}}\big(|C_\eta(0)|>1\big) \nonumber\\
& =&a(1-\rho)\alpha(\rho)\beta(\rho)+\big(1-(1-\rho)\alpha(\rho)\beta(\rho)-(1-\rho)\big) \nonumber\\
&=&
(a-1)(1-\rho)\alpha(\rho)\beta(\rho)+\rho. 
\end{eqnarray}
When $a=1$, we recover the SEP with $D(\rho)=G'(\rho)=1$. When $a\neq 1$, the parameters $\alpha(\rho)$ and $\beta(\rho)$ are functions of $\rho$ obtained by inverting Eqs.~\eqref{dainvertire}:
\begin{subequations}
\begin{align}
\label{alpha}
\alpha(\rho) &=\frac{-1+\sqrt{1+4(a-1)\rho(1-\rho)}}{2(a-1)(1-\rho)}\,,\\
\label{beta}
\beta(\rho) &=\frac{-1+\sqrt{1+4(a-1)\rho(1-\rho)}}{2(a-1)\rho}\,.
\end{align}
\end{subequations}
The diffusion coefficient is then computed as $D(\rho)=G'(\rho)$ to yield
\begin{equation}
\label{D(a)}
D(\rho) = \frac{2-1/\rho}{\sqrt{1+4(a-1)\rho(1-\rho)}}-\frac{1-\sqrt{1+4(a-1)\rho(1-\rho)}}{2(a-1)\rho^2}\,.
\end{equation}

\begin{figure}
\begin{center}
 \includegraphics[width=99mm,height=80mm]{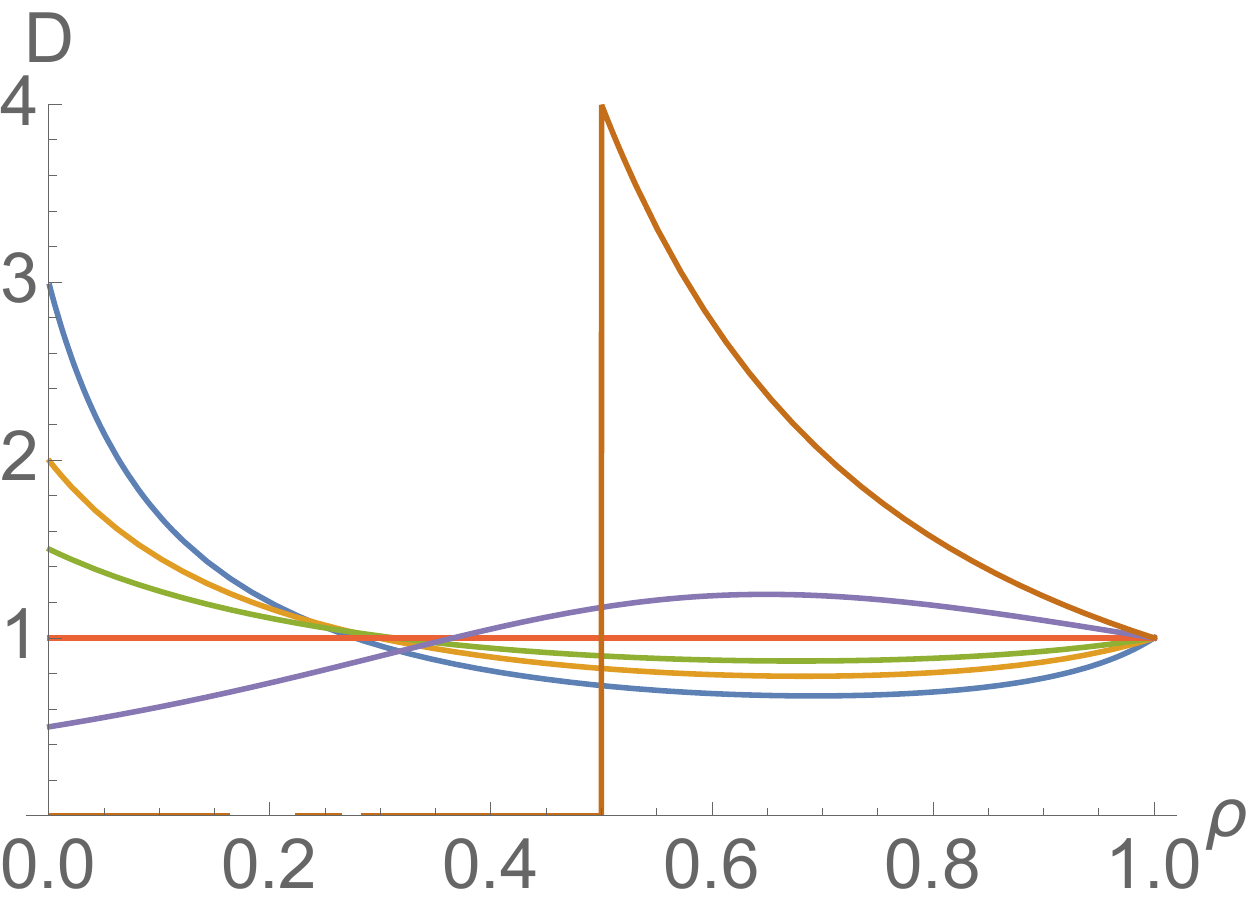}
 \caption{The diffusion coefficient $D(\rho)$ given by Eq.~\eqref{D(a)}. The definition \eqref{dubai} leads to the universal value $D(1)=1$ at the maximal density; at the minimal density $D(0)=a$. The curves corresponding to $a=0,\frac{1}{2},1,\frac{3}{2},2,3$ (they cross $\rho=0$ from bottom to top) are shown.}
\label{fig:D_AA}
\end{center}
\end{figure}

The diffusion coefficient is positive when $a>0$, while for $a=0$ we have
\begin{equation}
\label{D:FEP}
D(\rho) =
\begin{cases}
0              & 0\leq \rho < \frac{1}{2}\\
\rho^{-2}   & \frac{1}{2} < \rho \leq 1
\end{cases}
\end{equation}
This is the diffusion coefficient of a facilitated exclusion process, namely an exclusion process in which a particle can hop to  an empty neighboring site only when a complementary neighboring site is occupied. The totally asymmetric version of this facilitated exclusion process has been mostly studied, see e.g. \cite{Alan10,BBC}. The non-hydrodynamic behavior, $D(\rho)=0$ when $\rho<\frac{1}{2}$, is obvious in the context of the facilitated exclusion process---particles eventually get separated by at least one vacancy and the evolution stops. The hydrodynamic part of Eq.~\eqref{D:FEP} is easy to derive due to the connection with repulsion process \cite{RP13}; essentially the same formula has appeared as a limiting case in the previous section, see \eqref{D:RP}; and it will reappear again in the next section, Eq.~\eqref{DRR}.

\section{Strong hard rods}

In this Section we employ an argument that uses boundary conditions different from the periodic ones.
The SP dynamics can be naturally generalized to extended objects. In one dimension, the only connected extended objects are hard rods. In the case of zero strength, hard rods undergo a simple exclusion process; this is perhaps the first lattice gas model originally proposed to mimic RNA transcription \cite{MGB68,MG69}. The dynamics of hard rods of length $L\geq 2$ is the simplest case in which strong particles and leap frog dynamics are different. In the basic version of the strong dynamics each rod jumps one lattice site to the left or right with equal unit rates pushing away the other rods. For example, if a few hard rods are joined together in a single cluster and the leftmost rod is moving to the right the final result is that all the mass is moving one step to the right. In the case of the leap frog dynamics the final result is instead that all the mass is moving to the right by $L$ steps. (More precisely, this happens in the case when the gap between the rightmost hard rod in the cluster and the next cluster on the right is at least $L$.) In more than one dimension several interesting strong dynamics on extended objects are possible. For example, it would be very interesting to consider strong versions of models of the type studied in Ref.~\cite{LT}.

We can obtain the transport coefficients for strong hard rods using an
approach similar to the one in Section \ref{ewa}. In particular we consider a SP dynamics
on a closed interval with $N$ sites with $M=\lfloor \alpha N\rfloor$ particles. We map each configuration
of this model into a configuration of a model of strong hard rods of length $L$ evolving
on a closed interval of length $(L-1) \lfloor \alpha N\rfloor +N$ and having $M= \lfloor \alpha N\rfloor$ rods. The mapping
is the natural one and consists in substituting particles with rods and empty sites with empty
sites. Graphically it corresponds to
\begin{align}\label{defgrr}
& \bullet \ \ \ \Longrightarrow\ \ \  \overbrace{\bullet\cdots\bullet}^{L~\text{rod}}\\
& \circ \ \ \ \Longrightarrow\ \ \  \circ
\end{align}
where we represent an $L$ rod by a sequence of $L$ $\bullet$.
We call $\eta$ a configuration of strong particles and $\eta_R$ the corresponding configuration
of hard rods. We emphasize that $\eta_R(i)=1$ if the site $i$ contains a $\bullet$ of a rod.
The mapping \eqref{defgrr} can be written in formulas as
\begin{equation}\label{incendio}
L\sum_{j=1}^i\eta(j)=\sum_{j=1}^{(L-1)\sum_{k=1}^i\eta(k)+i}\eta_R(j)\,.
\end{equation}
When $\eta$ evolves according to SP dynamics then $\eta_R$ evolves according
to strong hard $L$-rods dynamics and the relationship \eqref{incendio} holds
for any time. Like in Section \ref{ewa} one finds that \eqref{incendio} corresponds
to the relationship
\begin{equation}\label{incendioM}
LF(y,t)=F_R\left((L-1)F(y,t)+y,t\right)
\end{equation}
between the distribution functions. The SP dynamics evolves macroscopically on the interval $[0,1]$ while
the strong hard $L$-rods dynamics evolves on the interval $[0,(L-1)\alpha+1]$. We use the same convention as in section \ref{ewa} for the arguments of functions having or not the lower index $R$. More precisely every time that we have a function with lower index $R$ the argument is $\left((L-1)F(y,t)+y,t\right)$ while every time that there is a function without the lower index $R$ the argument is simply $(y,t)$. Using the zero flux boundary conditions we arrive at
\begin{equation}
\label{bbb}
\partial_tF=D(\rho)\nabla \rho, \qquad \partial_tF_R=D_R(\rho_R)\nabla \rho_R\,.
\end{equation}
Differentiating \eqref{incendioM} once with respect to $t$ and once and twice with respect to $y$  we obtain
\begin{equation}\label{bbbb}
\left\{
\begin{array}{l}
L\partial_tF=\partial_tF_R+\rho_R(L-1)\partial_tF\,,\\
L\rho=\rho_R\big((L-1)\rho+1\big)\,,\\
L\nabla \rho=\big((L-1)\rho+1\big)^2\nabla\rho_R+\rho_R(L-1)\nabla\rho\,.
\end{array}
\right.
\end{equation}
Using \eqref{bbb} and \eqref{bbbb} we obtain
\begin{equation}
\label{sb}
D_R(\rho_R)=\frac{L^2}{[L-(L-1)\rho_R]^2}\, D\left(\frac{\rho_R}{L-(L-1)\rho_R}\right)\,.
\end{equation}
This relation is true for any SP dynamics with arbitrary jump rates.

For the 0-SP, i.e., for the simple exclusion process, we have $D=1$ which in conjunction with \eqref{sb} yield
the diffusion coefficient for a lattice gas of $L$-rods interacting by exclusion:
\begin{equation}
D_R(\rho_R)=\frac{L^2}{[L-(L-1)\rho_R]^2}\,.
\end{equation}
For the basic SP  the diffusion coefficient is given by \eqref{diffSP} which is combined with \eqref{sb} to give
\begin{equation}
D_R(\rho_R)=\frac{L+(2-L)\rho_R}{L(1-\rho_R)^3}\,.
\end{equation}

For $L=1$ we recover the diffusion coefficient of the SP, while in the $L\to\infty$ we arrive at a neat expression
\begin{equation}
\label{DRR}
D_R(\rho_R) = \frac{1}{(1-\rho_R)^2}\,.
\end{equation}
Interestingly, the same expression describes the diffusion coefficient in a process studied in section \ref{ewa}, see Eq.~\eqref{D:RP}, and also in the repulsion process \cite{RP13}.

\section{SP with relaxed exclusion}

We consider the situation  on which on each single site of the lattice there can be at most $M\geq 2$ particles.\footnote{Before we denoted by $M$ the total  number of particles; in this final section $M$ is the maximum number of particles that can occupy a single site.} This means that the state space is $\{0,1,\dots ,M\}^{\Lambda_N}$. The generalization of the exclusion process to this situation gives models that are not gradient \cite{KLO,KLO2}. Also in the case of SP the direct generalization of almost any previous model does not give a gradient model. In such generalized models the clusters of particles have a richer structure and are not automatically symmetric by inversion of the space so that the basic identity \eqref{curclu} is in general not satisfied. In this situation by a cluster we mean a subset of the graph contained between two consecutive empty sites. Several different gradient dynamics can be constructed but it is difficult to satisfy also the reversibility condition with respect to a given probability measure.

One exception is the extension to this situation of a special model among the ones discussed in section \ref{sec:DC}. (A similar restriction on the rates appears also for lozenge
tiling Glauber dynamics \cite{LT}.)
Given $i\in \Lambda_N$ we define
\begin{equation}\label{l+M}
d^{\pm}_i(\eta):=\inf\left\{n \geq 0\,:\, \eta(i\pm n)<M\right\}
\end{equation}
which is the distance to the left or to the right from site $i$ to the nearest not fully occupied site. (The root site $i$ is included, so the distance can be equal to zero.)
We also define
\begin{equation}\label{l+M0}
\hat d^{\pm}_i(\eta):=\inf\left\{n > 0\,:\, \eta(i\pm n)<M\right\}\,,
\end{equation}
that is similar to the above definition with the difference that $\hat d^\pm$ cannot be zero.
The elementary jumps of the dynamics are related to the elementary operators $\left[\sigma^{\pm}_i\eta\right]$ defined as follows.
If $\eta(i)=0$ or $\eta(j)=M$ for any $j\neq i$ then $\left[\sigma^{\pm}_i\eta\right]=\eta$. Otherwise we define
\begin{equation}\label{op++}
\left[\sigma^{\pm}_i\eta\right](j):=\left\{
\begin{array}{ll}
\eta(j)-1 & \textrm{if} \ j=i\,,\\
\eta(j)+1 & \textrm{if} \ j=i\pm\hat d^\pm_i\,,\\
\eta(j)     & \textrm{otherwise}\,.
\end{array}
\right.
\end{equation}
The flows associated to these elementary transformations are given by
$$
q^+_i(l,m):=\left\{
\begin{array}{ll}
1 & \textrm{if}\ m=l+1 \ \textrm{and}\  i\leq l\leq i+\hat d^+_i-1\\
0 & \textrm{otherwise}
\end{array}
\right.
$$
and likewise for jumps in the negative direction. The generator of the dynamics has the form
\begin{equation}
\label{gen-strong-k-M}
\mathcal L_N^{M,r}f(\eta):=\sum_{i\in \Lambda_N}\sum_{s=\pm}g(\eta(i))r\left(||q^{s}_i||\right)\big[f\big(\sigma^{s}_{i}\eta\big)-f\big(\eta\big)\big]\,.
\end{equation}
This model is gradient if $r(k)=k^{-1}$, and this is true for any function $g:\mathbb N\to \mathbb R^+$ such that $g(0)=0$.
This is a model of strong particles but it is easier described in terms of leap frog particles. The mechanism is the following. Starting from each site $i$ occupied by at least one particle, one particle jumps to the nearest free position at distance $n$ to the right with rate $g(\eta(i))n^{-1}$.  The rates of the jumps to the left are similarly defined. In terms of strong particles this corresponds to have strong particles that jump with a rate inversely  proportional to the number of particles that have to be leaped and directly proportional to the value of $g$ associated to the site from which particle jumps.

The dynamics is reversible with respect to the product measure on $\{0,1,\dots ,M\}^{\Lambda_N}$ defined by
\begin{equation}
\label{mup}
\mu_{M}^\lambda(\eta)=\prod_{i\in \Lambda_N}\frac{\lambda^{\eta(i)}}{Z_M(\lambda)g(\eta(i))!}\,.
\end{equation}
Here $\lambda$ is a parameter, and we use the shorthand notation
$$
g(k)!:=g(k)g(k-1)\dots g(1)
$$
with $g(0)!=1$ and denote by
\begin{equation}
\label{zebra}
Z_M(\lambda):=\sum_{k=0}^M\frac{\lambda^k}{g(k)!}
\end{equation}
a normalization constant. The model is defined by the values of $g$ on integers $\leq M$. Since we are discussing
also the limit $M\to +\infty$, we fix the function $g$ in such a way that
$$
Z_\infty(\lambda):=\sum_{k=0}^\infty\frac{\lambda^k}{g(k)!}
$$
is convergent for any $\lambda$ and $\lim_{\lambda\to +\infty}Z_\infty(\lambda)=+\infty$.

The reversibility is established by noting that if a jump is allowed the reversed one is also allowed, and by checking
all the possibilities concerning the number of particles in the two sites between which the particle is jumping and verifying that the detailed balance condition is satisfied.

The gradient condition is satisfied because with this choice of the rates the condition \eqref{curclu} is satisfied.
This is due to the fact that (recall that we consider the case $r(k)=k^{-1}$)
\begin{equation}\label{cud}
r(||q^+_i||)||q^+_i||=r(||q^-_i||)||q^-_i||=1
\end{equation}
for any $i$ and for any configuration so that 
\begin{equation}\label{curcluM}
\sum_{i: \{i,i+1\}\cap C\neq \emptyset}j_\eta(i,i+1)=\sum_{i\in C}g(\eta(i))
\left(r(||q^+_i||)||q^+_i||-r(||q^-_i||)||q^-_i||\right)=0\,.
\end{equation}
Clusters of particles are identified as in the case of models satisfying an exclusion rule but looking just at the presence of mass. More precisely, since we are in one dimension, the clusters of particles are the connected components of the graph obtained removing empty sites and the associated edges. We will introduce later another notion of clusters.
We call
\begin{equation}\label{lar}
\rho_M[\lambda]:=\frac{1}{Z_M(\lambda)}\sum_{k=0}^M\frac{k\lambda^k}{g(k)!}=\lambda \frac{d}{d \lambda}
\left(\log Z_M(\lambda)\right)
\end{equation}
the density of particles associated to the value $\lambda$ of the chemical potential and $\lambda_M[\rho]$ its inverse function. For $M=2$ this function can be explicitly computed:
\begin{equation}
\label{lac}
\lambda_2[\rho]=\frac{\sqrt{(1-\rho)^2 g^2(2)+4\rho(2-\rho)g(2)g(1)}+(\rho-1)g(2)}{2(2-\rho)}
\end{equation}
with density varying in the range $\rho\in (0,2)$.

We define a configuration $\tilde\eta\in \{0,1\}^{\Lambda_N}$ by $\tilde\eta(i)=1$ if $\eta(i)=M$ and zero otherwise. This means that the configuration $\tilde\eta$ identifies the sites that are full. We call $\tilde C$ the clusters associated to the configuration $\tilde\eta$. Given a cluster of particles $C$ it can contain none or several disjoints clusters $\tilde C_k$. We can split the current flowing across the cluster $C$ into a part $j^1$ coming from jumps of particles belonging to the clusters $\tilde C_k$ and a part $j^2$ coming from jumps of particles that do not belong to the clusters $\tilde C_k$. Both currents are of gradient type. Accordingly the diffusion coefficient will be computed as the sum of two parts, $D_M(\rho)=D_M^1(\rho)+D_M^2(\rho)$, coming respectively by the two components of the current.

The current $j^1$ is computed as follows. Consider a cluster $\tilde C_k$ and observe the current generated exclusively by the particles belonging to this cluster. This current is the same of that of a SP model with particles only in $\tilde C_k$ and having a rate of jump $g(M)k^{-1}$ for a jump moving $k$ particles. As we already discussed, this current is gradient; moreover, $j^1_\eta(i,i+1)=\tau_i h^1(\eta)-\tau_{i+1} h^1(\eta)$ with
$$
h^1(\eta)= H(d_0^-+d_0^+)- H(d_0^+)- H(d_0^-)
$$
where the function $H$ is defined by \eqref{H:def} and specialized to rates $r(k)=g(M)k^{-1}$. Note that under the measure $\mu_{M}^{\lambda[\rho]}$, the quantities $d_i^\pm$ are geometric random variables with
parameter $\frac{\lambda_M[\rho]^M}{Z_M(\lambda_M[\rho])g(M)!}$ that is the probability of having $M$ particles in one
single site. We can apply then the results of section \ref{sec:DC} obtaining
\begin{equation}\label{euna}
G_{M}^1[\rho]:=\mathbb E_{\mu_{M}^{\lambda[\rho]}}\left(h^1(\eta)\right)= \frac{g(M)\lambda_M[\rho]^M}{g(M)!Z_M(\lambda_M[\rho])-\lambda_M[\rho]^M}
\end{equation}
Observe that when $M\to +\infty$ we have $Z_M(\lambda)\to Z_\infty(\lambda)$ and $\lambda_M[\rho]\to \lambda_\infty[\rho]$.
Our assumptions imply that $\frac{\lambda^M}{g(M)!}\to 0$ for any $\lambda$ when $M\to +\infty$.
We deduce that the right-hand side of \eqref{euna} converges to $0$ and the contribution to the diffusion coefficient 
$$
D_{M}^1(\rho)=\frac{d G_{M}^1(\rho)}{d \rho}
$$
also converges to zero when $M\to +\infty$. Since the total current $j$ is gradient and $j^1$ is gradient then also $j^2$ that is obtained by the contributions of jumps of particles that belong to sites not completely full is also gradient.  Let $m=m(\eta)$ be the first empty site  on the left of the origin. This is the left boundary of the cluster of the origin. We define
\begin{equation}\label{mmm}
h^2(\eta)=-\sum_{i=m(\eta)}^{-1}j^2_\eta(i,i+1)\,.
\end{equation}
Using \eqref{cud} we find that the contribution to the total current coming from each single particle is zero. From this and a direct computation we get the following. If $\eta(0)<M$ there is only one contribution surviving in \eqref{mmm} coming from the jumps to the left of particles in $0$. If instead $\eta(0)=M$ then there are 3 contributions surviving. One positive coming from jumps to the left
of particles in $d_0^+$, one positive coming from the jumps to the left of particles in $-d_0^-$ and one negative coming from jumps to the right of particles in $-d_0^-$. After long but straightforward calculation one can express the contribution of all terms in a unified form valid for any case
$$
h^2(\eta)=g\left(\eta\left(-d_0^-\right)\right)\frac{d_{1}^++1}{d_0^-+d_{1}^++1}+
g\left(\eta\left(d_0^+\right)\right)\frac{d_{0}^-}{d_0^-+d_{1}^++1}
$$
Observe that
$$\mathbb E_{\mu_{M}^{\lambda[\rho]}}\left(g\left(\eta\left(d_0^+\right)\right)\frac{d_{0}^-}{d_0^-+d_{1}^++1}\right)=\mathbb E_{\mu_{M}^{\lambda[\rho]}}\left(g\left(\eta\left(-d_0^-\right)\right)\frac{d_{0}^-}{d_0^-+d_{1}^++1}\right)
$$
because under the measure $\mu_{M}^{\lambda[\rho]}$ the random variables
$\eta\left(d_0^+\right), \eta\left(-d_0^-\right), d_0^-, d_{1}^+$ are all independent and moreover $\eta\left(d_0^+\right)$ and $\eta\left(-d_0^-\right)$ have the same distribution. Since $\eta(-d_0^-)\in \left\{0,1,\dots ,M-1\right\}$ and moreover
$$
\mathbb P_{\mu_M^{\lambda[\rho]}}\left(\eta(-d_0^-)=j\right)=\frac{\lambda[\rho]^j}{Z_{M-1}(\lambda[\rho])g(j)!}
$$
we deduce
$$
G_M^2(\rho)=\mathbb E_{\mu_{M}^{\lambda[\rho]}}\left(h_2(\eta)\right)=\mathbb E_{\mu_{M}^{\lambda[\rho]}}\left(g\left(\eta\left(-d_0^-\right)\right)\right)
$$
that is equal to
\begin{equation}
\label{g2}
G_{M}^2(\rho)=\sum_{j=0}^{M-1}\frac{g(j)\lambda[\rho]^j}{Z_{M-1}(\lambda_M[\rho])g(j)!}=
\frac{\lambda_M[\rho]Z_{M-2}(\lambda_M[\rho])}{Z_{M-1}(\lambda_M[\rho])}\,.
\end{equation}
Therefore we finally obtain
$$
D_M(\rho)=\frac{d\left[G_{M}^1(\rho)+G_{M}^2(\rho)\right]}{d \rho}\,,
$$
where $G^1_M(\rho)$ is given by \eqref{euna} while $G^2_M(\rho)$ is given by \eqref{g2}.
In the case $M=2$, using \eqref{lac} it is possible to have a completely explicit form of $D_2(\rho)$.

Note that $G_{M,2}(\rho)\to \lambda_\infty[\rho]$ when $M\to \infty$. Since we have already shown
that $D_M^1$ is converging to zero for large $M$, we obtain
$$
\lim_{M\to \infty}D_M(\rho)= \frac{d\left(\lambda_\infty[\rho]\right)}{d\rho}
$$
that is the diffusion coefficient of a zero range dynamics with rates of jump determined by the function $g$.

\bigskip
\bigskip

{\bf Acknowledgements.}
We thank the Galileo Galilei Institute for Theoretical Physics for excellent working conditions and the INFN for partial support in the earlier stage of this work. D. G.  thanks   the Institute Henri Poincar\'e   for the hospitality and  the support during the trimester  ``Stochastic Dynamics Out of Equilibrium''.

\end{document}